# A New Class of Indicators for the Model Selection of Scaling Laws in Nuclear Fusion

I.Lupelli1, A.Murari2, P.Gaudio5, M.Gelfusa5, D.Mazon3, J.Vega4

*1)EURATOM/CCFE Fusion Association, Culham Science Centre, OX14 3DB Abingdon (UK)*
*2) Consorzio RFX-Associazione EURATOM ENEA per la Fusione, I-35127 Padova, Italy.*
*3) Association EURATOM-CEA, CEA Cadarache DSM/IRFM, 13108 Saint-Paul-lez-Durance, France.*
*4) Asociación EURATOM-CIEMAT para Fusión, CIEMAT, Madrid, Spain.*
5)Univerisity of Rome Tor Vergata, Via del Politecnico 1 00133 Roma, Italy

The development of computationally efficient model selection strategies represents an important problem facing the analysis of Nuclear Fusion experimental data, in particular in the field of scaling laws for the extrapolation to future machines, and image processing. In this paper, a new model selection indicator, named Model Falsification Criterion (MFC), will be presented and applied to the problem of choosing the most generalizable scaling laws for the power threshold ($P_{Thresh}$) to access the H-mode of confinement in Tokamaks. The proposed indicator is based on the properties of the model residuals, their entropy and an implementation of the data falsification principle. The model selection ability of the proposed criterion will be demonstrated in comparison with the most widely used frequentist (Akaike Information Criterion) and bayesian (Bayesian Information Criterion) indicators.

## 1. Introduction

Karl Popper argued in favour of a simple way to determine the scientific validity of a theory: the falsification principle. This principle states that, in order for a theory to be deemed scientific, there must exist an empirical way of showing that it is false. A model is a quantitative formulation of a theory, which provides falsifiable predictions of phenomena. Testing the reliability of models and scaling laws is very important during the conceptual study or design phase of a new experiment such as ITER. For example, the $P_{Thresh}$[1][2] to access the H-mode of confinement in Tokamaks represents a critical engineering parameter for the design of the additional heating systems and for the extrapolations of operative scenarios. A goodness-of-fit measure (GOF) is usually used to measure the quality of quantitative models. What is measured is how much a model's predictions deviate from the observed data [3,4,5]. The model that provides the best fit (i.e. the smallest deviation) is favoured. Unfortunately one of the biggest challenges faced by Nuclear Fusion scientists is that experimental signals are noisy, or in any case affected by large error bars. Errors arise from several sources, such as the imprecision of measurement tools, electronic noise etc. Noisy data make the simply GOF measure by itself a poor method of model selection. A typical GOF measure such as the Root Mean Squared Error (RMSE), for example, is insensitive to the number of the explanatory variables, dimensionality of the dataset and different sources of variation in the data. This could result in the selection of a model that overfits the data, which may not be the model that best approximates the process under study. The preferred solution has been to redefine the problem as one of assessing how well a model's fit to one data sample generalizes to future samples generated by that same process [6]. Early measures of generalizability are the Akaike information criterion (AIC) [7] and Bayesian information criterion (BIC) [7]. In this approach a good fit is a necessary but not sufficient condition for a model because more complex models are penalised (as a form of parsimony). In this paper, an original model selection indicator, named Model Falsification Criterion (MFC), will be presented. The new proposed indicator is based on the properties of the model residuals, their entropy and an implementation of the data falsification principle. The model selection ability of the proposed indicator will be demonstrated by comparison with the AIC and BIC by using first two datasets: a low dimensionality real-life dataset and a synthetic high dimensionality dataset. Moreover the indicator will be applied to the 2010 version of the ITPA database (IGDBTHv6b)[1,8,9] in order to derive the most generalizable scaling laws for the $P_{Thresh}$ as function of a limited number of macroscopic quantities.

## 2. The Model Falsification Criterion

### 2.1 Theoretical background

The present version of the criterion, name Model Falsification Criterion (MFC), is conceived as a measure of generalizability of a model. The generalizability is the ability of a model to fit all data samples generated by the same process under study, not just the currently observed samples. Generalizability is estimated by combining a model's GOF with a measure of its complexity. A complex model, with many parameters, could in principle be more adequate to interpret the data generated by a complex system. Therefore, in order to implement some form of parsimony, the main inspiration in the formulation of the MFC criterion has been the falsification principle more than the Occam's Razor. A model is to be preferred when a small error in its parameters does not result in a great change in its estimates. In this context, the falsification principle is

---
*author's email: ivan.lupelli@uniroma2.it*

therefore interpreted and translated in terms of the robustness of the model. The principle of parsimony, to increase the generalization capability of the indicator, is therefore implemented in such a way as to penalize not so much the simple number of parameters but more the repercussions on the final estimates of small errors in the model parameters. In mathematical terms, the main elements of the MFC criterion are based on the properties of the model residuals (r), their entropy H(|r|) and the robustness of the model estimates against the variations in the j-th explanatory variables included in the model (falsification principle). The value of the MFC for a model under study is:

$$\text{(Eq. 1) MFC}(r,k,n) = \overbrace{\left(k_{corr} + 2k + \frac{\Sigma|r|}{H(|r|)}\right)}^{MFC_1} + \overbrace{\sum_{j=1}^{R} k_{corr} + 2k + \frac{nk_{corr}}{2k^2}\frac{\Sigma|r_j|}{H(|r_j|)}}^{MFC_2}$$

where r represents the model's residuals vector, H(|r|) estimates the Shannon entropy [9] of the r from the corresponding observed counts, k is the number of model parameters (i.e the number of estimated coefficient of the model plus the variance), $r_j$ represents the value of model residuals vector, calculated after varying inside the error bars each explanatory variable from a set of R candidate variables $x_j$ ($x_j$={$x_1,x_2,….,x_R$}) included in the model one at the time. $k_{corr}$=2k(k+1)/(n-k-1) is the corrected number of model parameters that takes into account the total number of observations n and the model parameters. In analogy with Information Theory [6] $k_{corr}$ provide a greater penalty for complex models. In Eq.1 the first part of the indicator, $MFC_1$, takes into account both the properties of the model residuals, how well the model fits the experimental data, and the model complexity. A good model should have a low sum of the residuals and a high entropy of the residuals keeping the model complexity (provided by $k_{corr}$ and k) as low as possible. The second part, $MFC_2$, assesses the robustness of the model against the falsification in explanatory variables. In the case of a robust model, the sum of the residuals should not increase significantly and the entropy of the residuals should not decrease significantly in the case of variations of its explanatory variables inside their error bars. $MFC_2$ is calculated introducing an error on each explanatory variable of the model one at the time. According to the proposed criterion, the most generalizable model is the one which presents the lowest value of the MFC indicator. The mathematical definition of the MFC criterion has to be compared with the ones of the AIC = 2k + nln(RSS/n) and BIC = nln(RSS/n-1) + kln(n) where *n* is the number of samples in the database, *k* the number of parameters in the model, RSS is the sum of squared residuals and ln indicates the natural logarithm.

## 3. Databases description

### 3.1 Test databases description

Dataset 1 [7] is a simple low-dimensionality experimental database consisting of 13 observations of 4 of candidate variables ($x_1$ to $x_4$) and 1 dependent variable (y). The true model is the model that includes just $x_1$ and $x_2$ as explanatory variables y=52.6+1.468$x_1$+0.662$x_2$ with $x_3$ and $x_4$ as spurious variables. This dataset has been considered in order to test the performance of the MFC criterion in a case when both AIC and BIC identify the correct and well established experimental model [7]. The Dataset 2 is a high-dimensionality synthetic database generated by a linear model with one response variable y and 12 candidate variables $x_1$ through $x_{12}$ (150 observations). The true model is y=$x_1$+$x_2$+$x_3$+$x_4$+$x_5$ and variables from $x_6$ through $x_{12}$ are considered spurious. The validity of the results has been tested with more complex linear and non-linear functional forms [7]. This dataset has been considered in order to test the performance of the MFC in presence of an high number of spurious variables in the set of candidate regressors. For R predictors, since each can be either included or omitted (two possibilities for each variable), there are $2^R$-1 (excluding the trivial solution with no variables) possible models for each dataset: 15 model for Dataset 1 and 4095 for Dataset 2. For both datasets the noise level introduced in the explanatory variables amounts to ±10% of their original value.

### 3.2 The ITPA International Multi-machine Database

This MFC indicator has been applied to the ITPA International Global Threshold Data Base v6b(IGDBTHv6b) of L to the H mode transitions [8] according to the selection criteria (SELEC2007), defined to extract ITER relevant discharges from the database. According to these criteria, discharges with single-null configurations, ion grad B drift towards the X point and deuterium as fuel are selected. On the other hand, plasmas with too low plasma density [1], too low safety factor at the 95% flux surface ($q_{95}$< 2.5), too large counter-NBI fraction (Pctr/$P_{NBI}$> 0.8), too small gaps between plasma surface and wall (d < 5 cm) are discarded. These criteria furthermore exclude transitions obtained in Ohmic conditions, since they are not relevant for ITER, and Electron Cyclotron only heated discharges, since this heating scheme, mainly used in small devices, regularly leads to high $P_{Thresh}$ values [1]. Also configurations with a plasma elongation lower than 1.2 have not been considered. To maximize the accuracy of the scalings, only the data belonging to an interval of 50 ms before the L-H transition has been included in this analysis; all together a total of 470 discharges from the main Tokamaks (JET, AUG, DIIID, CMOD) present all the required quantities in the aforementioned interval and can be used in the analysis. Previous studies have shown that, in terms of macroscopic quantities measured on all major devices [1][2][8], $P_{Thresh}$ mainly depends on the plasma line integrated density ($n_{e20}$[$10^{20}m^{-3}$]), the strength of the toroidal field (Bt[T]), the plasma surface area (S[$m^2$]) major (R[m]) and minor radius (a[m]), elongation (k), triangularity (δ), the plasma current Ip [MA] and $q_{95}$ (the safety factor at 95% of the plasma radius). The operational range covered by this set of discharges is 1.29 < $B_t$[T] < 5.37, 0.20 < $n_{e20}$[$10^{20}$ $m^{-3}$] < 1.19, 0.67 < R[m] < 2.92, 0.216 < a[m] < 1.03, 7.32 <

**Figure 3** – Histogram of Residuals for the models proposed by different criteria (AIC,BIC,MFC) using the whole multi-machine ITPA database. The residuals of each model have also been fitted with a normal distribution, whose mean and standard deviation (StDev) values are reported.

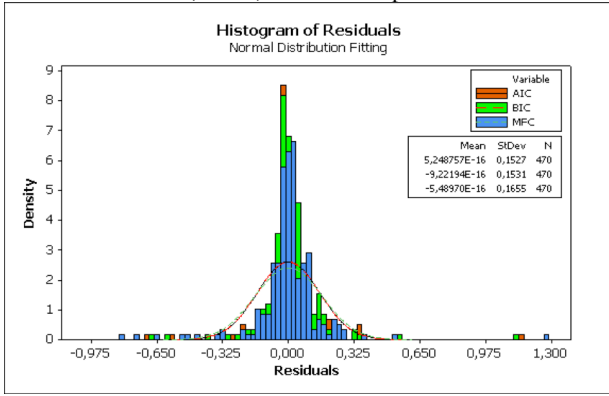

$S[m^2] < 174.10$, $0.0696 < \delta < 0.508$, $1.56 < k < 2.04$, $2.51 < q_{95} < 6.78$, $0.54 < I_p[MA] < 3.22$ and $0.831 < P_{Thresh}[MW] < 6.466$. All these quantities are routinely available in all the major Tokamaks, providing enough data for a sound statistical analysis (for the single machine analysis just JET, DIID and AUG have been considered). The noise level for each variable has been assumed as ± 10%. The scaling laws considered in this paper are of the form of power-law monomials

(Eq.2) $P_{Thresh} = \beta_0 a^{\beta_1} R^{\beta_2} S^{\beta_3} \delta^{\beta_4} k^{\beta_5} B_t^{\beta_6} n_{e20}^{\beta_7} q_{95}^{\beta_8} I_p^{\beta_9}$

## 4. Results

The proposed most generalizable models provided by the different criteria (AIC, BIC, and the proposed MFC), are reported in Table 1. For the low dimensionality experimental dataset (Dataset 1) the MFC is able to find the correct true model of the process under study (4 parameters and two variables $x_1, x_2$). This means that the MFC has at least the same selection power of the classic indicators for databases of low dimensionality experimental problems. For the high dimensionality dataset (Dataset 2), it appears very clearly that the MFC outperforms the traditional AIC and BIC criteria. In Figure 1 the relationship between GOF (RMSE) and the generalizability (AIC, BIC and MFC value) as a function of model complexity (k) is shown. It is clear that the MFC penalizes more the complex models with respect to the AIC and BIC criteria, and also is able to find the correct model in a high dimensionality dataset. This is a very important point, especially given the high dimensionality and noise level of the data in Nuclear Fusion, when it comes to important applications such as the identification of multivariate scaling laws or large datasets of image processing data. It has also been checked that the MFC preserves its advantages up to a noise of 20% of the signal level. The result of MFC applied to the problem of choosing the most generalizable scaling laws for the $P_{Thresh}$ to access the H-mode of confinement in Tokamaks are showed in Table 2. Both AIC and BIC criteria identify more complex models than the MFC. The results provided by the MFC in Figure 2, show again the ability of the indicator to select the most generalizable and less complex model for the L-H transition [2] (dependence of $P_{Thresh}$ from the geometrical quantities, magnetic field and electron density). For the reader convenience, the scaling laws are explicitly reported in the following:

(Eq. 3) $P_{Thresh} (ITPA) = 0.686_{0.678}^{0.694} a^{0.961_{0.958}^{0.964}} R^{1.072_{1.071}^{1.074}} S^{-0.00009_{-0017}^{0.0015}} B_t^{0.7318_{0.7315}^{0.732}} n_{e20}^{0.749_{0.748}^{0.794}}$

(Eq. 4) $P_{Thresh} (JET) = 1.781_{1.676}^{1.887} B_t^{0.763_{0.659}^{0.868}} n_{e20}^{0.774_{0.695}^{0.854}}$

(Eq. 5) $P_{Thresh} (DIIID) = 0.773_{0.747}^{0.799} B_t^{0.743_{0.712}^{0.773}} n_{e20}^{0.740_{0.722}^{0.758}}$

(Eq. 6) $P_{Thresh} (AUG) = 0.540_{0.531}^{0.550} B_t^{0.736_{0.728}^{0.744}} n_{e20}^{0.723_{0.716}^{0.729}}$

Figure 3 reports the histograms of the residuals for the models proposed by different criteria (AIC, BIC, MFC) using the whole multi-machine ITPA database. The pdf of the model selected by the MFC criterion is only slightly better than the ones of the other two models (indentified by AIC and BIC) but it includes a significantly lower number of variable (lower complexity). Particularly interesting are also the results for the individual machines (the results for JET machine are shown in figure 2), for which the MFC is the only criterion capable of identifying the fact that the geometric variables are not good regressors, since they do not vary sufficiently on a single machine. Of course this does not imply that a simple use of the MFC can substitute more involved analysis of the datasets under investigation (ANOVA, collinearity analysis etc.) but certainly shows a competitive advantage of the criterion compared to the more traditional ones.

**Table 1** – Models proposed by different criteria for the Dataset 1 and Dataset 2.

| Dataset: | Spurious variables | Variables included in the true model | Variables included in the model according to different criteria | | |
|---|---|---|---|---|---|
| | | | AIC | BIC | MFC |
| Dataset 1 (13 observations) | $x_3, x_4$ | $x_1, x_2$ | $x_1, x_2$ | $x_1, x_2$ | $x_1, x_2$ |
| Dataset 2 (150 observations) | $x_6$ to $x_{12}$ | $x_1$ to $x_5$ | $x_1, x_2, x_3, x_4, x_5, x_9$ | $x_2, x_3, x_4, x_5$ | $x_1, x_2, x_3, x_4, x_5$ |

**Table 2** – Models proposed by different criteria for the multi-machine ITPA database and for the single machines (JET,DIID,AUG)

| | Variables included in the model according to different criteria | | |
|---|---|---|---|
| Data | AIC | BIC | MFC |
| ITPA | $a, R, S, \delta, k, B_t, n_{e20}, q_{95}, I_p$ | $a, R, S, \delta, k, B_t, n_{e20}$ | $a, R, S, B_t, n_{e20}$ |
| JET | $R, \delta, B_t, n_{e20}, I_p$ | $R, B_t, n_{e20}, I_p$ | $B_t, n_{e20}$ |
| DIID | $S, \delta, k, B_t, n_{e20}, q_{95}, I_p$ | $S, k, B_t, n_{e20}$ | $B_t, n_{e20}$ |
| AUG | $a, R, S, \delta, k, B_t, n_{e20}, q_{95}, I_p$ | $a, R, S, \delta, k, B_t, n_{e20}, q_{95}$ | $B_t, n_{e20}$ |

**Figure 1** –Relationship between goodness of fit, expressed in terms of Root Mean Squared Error (RMSE) and generalizability measure, expressed in term of AIC BIC and MFC as a function of model complexity. Results for Dataset 1 and 2 are reported. The vertical lines correspond to the best model identified by the various indicators.

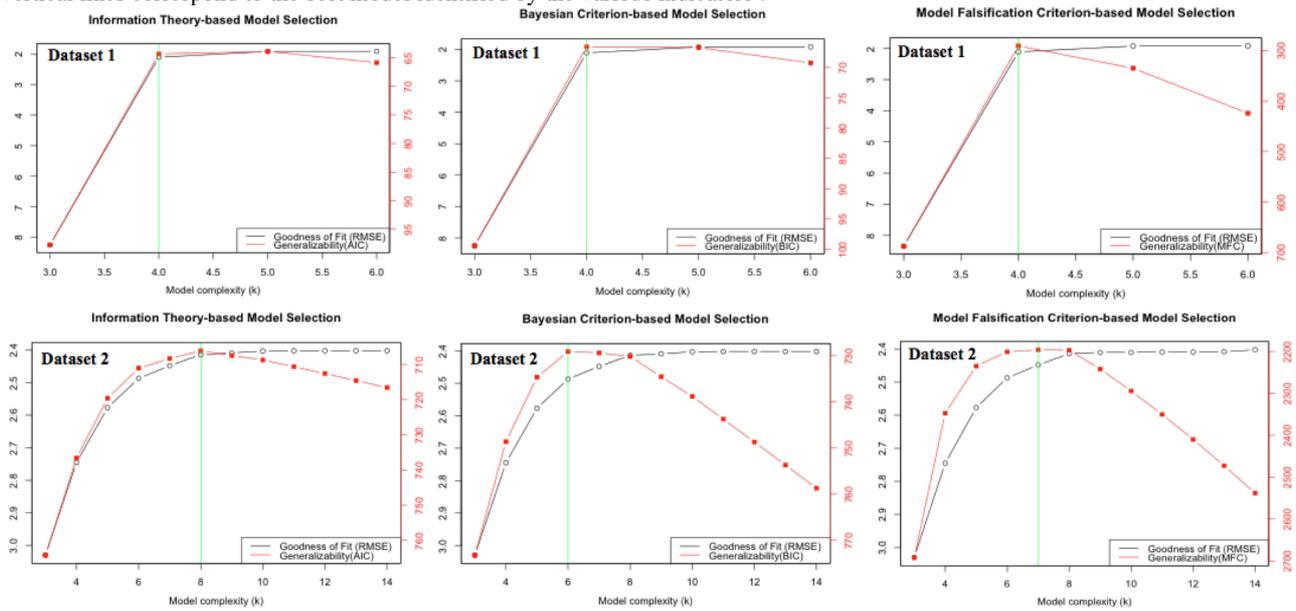

With regard to future developments, asymptotic properties of the MFC, its application in conjunction with the symbolic regression tools and the extension to image processing applications will be considered.

**Figure 2** Relationship between goodness of fit, expressed in terms of Root Mean Squared Error (RMSE) and generalizability measure, expressed in term of AIC BIC and MFC as a function of model complexity. Results for multi-machine ITPA (10% noise level) database and an individual machine (JET) are reported (10% noise level). The green vertical lines correspond to the best model identified by the various indicators. A 10% of noise has been considered for each variable.

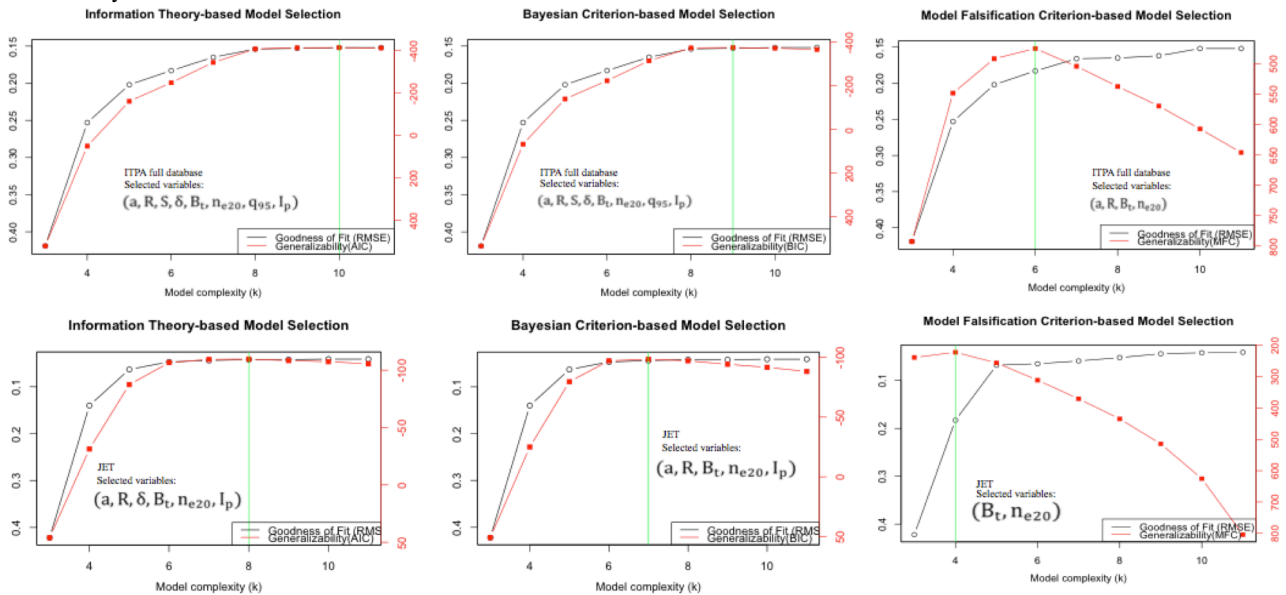